\documentclass[conference]{IEEEtran}
\IEEEoverridecommandlockouts

\usepackage{cite}
\usepackage{amsmath,amssymb,amsfonts}
\usepackage{algorithmic}
\usepackage{graphicx}
\usepackage{textcomp}
\usepackage{xcolor}
\usepackage{url}
\usepackage{algorithm}
\usepackage{algorithmic}
\usepackage{bm}
\usepackage{tikz}
\usetikzlibrary{positioning}
\usepackage{tkz-graph}
\usepackage{pgfplots}
\usepackage{listofitems}
\usepackage{flushend}
\usetikzlibrary{math, calc, positioning, shapes, backgrounds, fit, arrows, decorations.pathreplacing}
\pgfplotsset{compat=1.18}

\DeclareMathOperator*{\argmin}{arg\,min}

\def\BibTeX{{\rm B\kern-.05em{\sc i\kern-.025em b}\kern-.08em
    T\kern-.1667em\lower.7ex\hbox{E}\kern-.125emX}}
\begin{document}

\title{Iterative Neural Rollback Chase-Pyndiah Decoding}

\author{
  \IEEEauthorblockN{Dmitry Artemasov, Oleg Nesterenkov, Kirill Andreev, Pavel Rybin, Alexey Frolov}\textcolor{white}{\tiny{,}}

  \IEEEauthorblockA{\textit{Center for Next Generation Wireless and IoT}}
  \IEEEauthorblockA{\textit{Skolkovo Institute of Science and Technology}\\
  Moscow, Russia \\
  \{d.artemasov, o.nesterenkov, k.andreev, p.rybin, al.frolov\}@skoltech.ru}

  \thanks{The research was carried at Skolkovo Institute of Science and Technology and supported by the Russian Science Foundation (project no. 23-11-00340), \protect\url{https://rscf.ru/en/project/23-11-00340/}}
}

\maketitle

\begin{abstract}
Iterative decoding is essential in modern communication systems, especially optical communications, where error-correcting codes such as turbo product codes (TPC) and staircase codes are widely employed. A key factor in achieving high error correction performance is the use of soft-decision decoding for component codes. However, implementing optimal maximum a posteriori (MAP) probability decoding for commonly used component codes, such as BCH and Polar codes, is computationally prohibitive. Instead, practical systems rely on approximations, with the Chase-Pyndiah algorithm being a widely used suboptimal method. TPC are more powerful than their component codes and begin to function effectively at low signal-to-noise ratios. Consequently, during the initial iterations, the component codes do not perform well and introduce errors in the extrinsic information updates. This phenomenon limits the performance of TPC. This paper proposes a neural network-aided rollback Chase-Pyndiah decoding method to address this issue. A transformer-based neural network identifies cases where extrinsic updates are likely to introduce errors, triggering a rollback mechanism which prevents the update and keeps the component code message intact. Our results demonstrate that a neural network with a relatively small number of parameters can effectively distinguish destructive updates and improve decoding performance. We evaluate the proposed approach using TPC with (256, 239) extended BCH component codes. We show that the proposed method enhances the bit error rate performance of Chase-Pyndiah $\bm{p=6}$ decoding, achieving a gain of approximately $\bm{0.145}$~dB in a TPC scheme with four full iterations, significantly outperforming conventional Chase $\bm{p=7}$ decoding.
\end{abstract}

\begin{IEEEkeywords}
channel decoding, product codes, deep neural networks, soft-output decoding, turbo product codes
\end{IEEEkeywords}

\section{Introduction}
\label{sec:intro}

Error-correcting coding is fundamental to modern communication systems, ensuring reliable data transmission across various channels. In high-speed wireless and optical networks, iterative soft-decision decoding techniques are extensively employed to achieve near-optimal performance while maintaining manageable computational complexity. Among these, turbo product codes (TPCs) have gained prominence due to their ability to enhance bit error rate (BER) performance significantly~\cite{Mukhtar2016}. A critical aspect of these systems is the soft-decision decoding of component codes, such as Bose–Chaudhuri–Hocquenghem (BCH) and Polar codes~\cite{Bose1960, Arikan2009}. However, implementing optimal maximum a posteriori (MAP) probability decoding for these codes is often computationally impractical.

To address this challenge, suboptimal yet computationally efficient methods like the Chase-Pyndiah algorithm are widely adopted~\cite{Pyndiah1998}. This algorithm employs the soft output approximation of the decoded message by taking into account the list of possible codewords (competitors) produced by the Chase algorithm for each of the component codewords~\cite{Chase1972}, facilitating performance improvements with each iteration. However, a notable limitation of this approach is the potential introduction of erroneous extrinsic information during updates, which can degrade decoding performance. This issue becomes particularly significant due to the nature of the iterative decoding schemes, in which such an error negatively affects the convergence.

Some recent studies have addressed the issue of the Chase-Pyndiah algorithm's computational complexity and performance. For instance, in works~\cite{Graf2021, Strashofer2023} authors derived a Chase-Pyndiah post-processing function based on the mutual information estimation aiming to improve the resulting soft-output distribution. In another paper~\cite{Yoon2020} authors introduced techniques to reduce the decoding complexity of turbo product codes: an algorithm that minimizes hard decision decoding operations and an early termination technique for undecodable blocks, effectively lowering computational complexity.

At the same time, several studies have investigated the application of machine learning (ML) methods for TPC decoding. For example, in~\cite{Artemasov2025}, the authors proposed a syndrome-based soft-input soft-output neural decoder, which attained MAP component code decoding performance within the TPC scheme. Other studies~\cite{TenBrink2023, Jamali2022} have explored the product autoencoder framework, in which both the encoder and decoder are initialized as trainable blocks and optimized jointly.

Building upon these advancements, we propose a neural network-aided rollback Chase-Pyndiah decoding method to mitigate the adverse effects of erroneous extrinsic information updates. Specifically, we employ a transformer-based neural network to identify instances where an extrinsic information update is more likely to introduce errors than improve decoding performance. When such instances are detected, a rollback mechanism is activated, preventing the update and preserving the integrity of the component code's message. This approach ensures that only beneficial updates are applied, thereby enhancing overall decoding reliability and convergence rate. Notably, the proposed method retains the parallelizable structure of TPC decoding. By incorporating a separate rollback decision block, it preserves the classical Chase-Pyndiah decoding algorithm architecture, ensuring easy integration into existing decoding solutions. In this paper, we evaluate the proposed approach using TPC with (256, 239) extended BCH (eBCH) component codes. We demonstrate that the proposed method enhances the BER performance of Chase-Pyndiah $p=6$ decoding, achieving a gain of approximately $0.145$~dB in a TPC scheme with four full iterations, significantly outperforming conventional Chase $p=7$ decoding.

The remainder of this paper is organized as follows: Section~\ref{sec:system-model} introduces the system model. Section~\ref{sec:chase-pyndiah-decoding} provides a brief introduction to the Chase-Pyndiah decoding algorithm and its limitations. Section~\ref{sec:neural-rollback} introduces the rollback mechanism and describes the proposed neural network architecture. Section~\ref{sec:experiments-results} defines the experimental setup, presents simulation results, and provides performance analysis. Finally, Section~\ref{sec:conclusion} concludes the paper and outlines potential directions for future research.

\section{System Model}
\label{sec:system-model}

\begin{figure}[tb]
\centering
\includegraphics{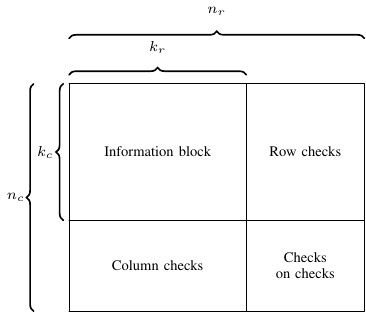}
\caption{Construction of turbo product code $\mathcal{P}=\mathcal{C}_c\otimes\mathcal{C}_r$.}
\label{fig:tpc-structure}
\end{figure}

Let us consider a communication system that employs the TPC scheme based on two component binary linear codes: $\mathcal{C}_c$ -- column component code, and $\mathcal{C}_r$ -- row component code, with lengths $n_c$, $n_r$, and dimensions $k_c$, $k_r$, respectively. Component codes $\mathcal{C}_c$ and $\mathcal{C}_r$ are defined by parity-check $\mathbf{H}_c$, $\mathbf{H}_r$ and generator $\mathbf{G}_c$, $\mathbf{G}_r$ matrices in systematic form. The product code $\mathcal{P}=\mathcal{C}_c\otimes\mathcal{C}_r$ is generated through the following steps (see~\figurename\ref{fig:tpc-structure}):
\begin{enumerate}
    \item Arranging $k_c\cdot k_r$ information bits into an array with $k_c$ rows and 
$k_r$ columns.
    \item Encoding each of the $k_c$ rows using code $\mathcal{C}_r$.
    \item Encoding each of the $n_r$ columns using code $\mathcal{C}_c$.
\end{enumerate}
The parameters of the constructed product code $\mathcal{P}$ are $(n_cn_r, k_ck_r)$, and the code rate is $R=(k_ck_r)/(n_cn_r)$.

After encoding the information sequence $\mathbf{U}\in\{0,1\}^{k_c\times k_r}$ into a TPC codeword $\mathbf{C}\in \{0,1\}^{n_c\times n_r}$, modulation is applied. In this paper, we employ binary phase-shift keying (BPSK), defined by the following mapping:
\begin{equation}
    \label{eq:modulation}
    \mathbf{X}_{i,j} = \tau(\mathbf{C}_{i,j}), \:\: \forall i\in [1,\dots , n_c], \: j\in [1,\dots , n_r],
\end{equation}
where $\tau:\{0, 1\} \rightarrow \{1, -1\}$.

Next, we consider transmission over a binary-input additive white Gaussian noise (AWGN) channel, where the receiver observes
\begin{equation}
    \mathbf{Y} = \mathbf{X} + \mathbf{Z}.
\end{equation}
Here $\mathbf{Z}$ is additive Gaussian noise matrix whose elements are i.i.d. sampled from the normal distribution with zero mean and $\sigma^2$ variance, i.e. $\mathbf{Z}_{ij}\overset{\text{i.i.d.}}{\sim}\mathcal{N}\left(0, \sigma^2\right)$. The impact of noise on the received signal is quantified by the signal-to-noise ratio (SNR), given by $E_s/N_0 = 1/2\sigma^2$. 

As is common in the literature \cite{Ryan2009}, the decoder input is represented by a matrix of log-likelihood ratios (LLRs), $\bm{\Gamma} \in\mathbb{R}^{n_c\times n_r}$, where each element is given by
\begin{multline}
    \bm{\Gamma}_{i,j} = \log\frac{p(\mathbf{Y}_{i,j} | \mathbf{C}_{i,j} = 0)}{p(\mathbf{Y}_{i,j} | \mathbf{C}_{i,j} = 1)} = \frac{2 \mathbf{Y}_{i,j}}{\sigma^2}, \\ \forall i \in [1,\dots, n_c], j \in [1,\dots, n_r],
\end{multline}
where $\log(\cdot)$ denotes the natural logarithm. 

In this paper, we evaluate the decoding performance using the bit error rate metric. Let $\mathbf{\hat{U}} \in \{0,1\}^{k_c\times k_r}$ denote the estimated information word. Then BER is defined as
\begin{equation}
P_b = \frac{1}{k_ck_r}\sum\nolimits_{i=1}^{k_c}\sum\nolimits_{j=1}^{k_r} \Pr[\mathbf{U}_{i,j} \neq \mathbf{\hat{U}}_{i,j}].
\end{equation}

\section{Chase-Pyndiah decoding}
\label{sec:chase-pyndiah-decoding}

We begin with a brief overview of TPC Chase-Pyndiah decoding. For a detailed description of the complete algorithm, we refer the reader to~\cite{Pyndiah1998}. The decoding process operates on the LLRs matrix, $\bm{\Gamma}$. The TPC decoder first decodes columns once with the Chase-Pyndiah algorithm, then rows, repeating this process for a predefined number of iterations, $N_{T}$.

Let us denote the a posteriori matrix at half-iteration ${t\in[1,\dots,2N_T]}$ as $\mathbf{L}_t$, initialized as $\mathbf{L}_0 = \bm{\Gamma}^\prime$. For odd $t$, the column message update is given by
\begin{equation}
    \label{eq:col-update}
    \mathbf{L}_t = \bm{\alpha}_t \mathbf{W}^{\prime c}_t + \bm{\Gamma}^\prime.
\end{equation}
Here $\bm{\alpha}_t$ is a weighting factor. $\mathbf{W}^{\prime c}_t$ represents the normalized extrinsic information matrix, scaled to have a mean of one. $\bm{\Gamma}^\prime$ denotes the normalized LLR matrix~\cite{Pyndiah1998}.

For even $t$, the row message update is performed in a similar manner
\begin{equation}
\label{eq:row-update}
    \mathbf{L}_t = \bm{\alpha}_t \mathbf{W}^{\prime r}_t + \bm{\Gamma}^\prime.
\end{equation}

The extrinsic information matrix $\mathbf{W}_t$ is derived from the a posteriori matrix $\mathbf{L}_{t-1}$ in two steps. First, the Chase~\cite{Chase1972} algorithm is applied to each component message $\bm{l}\subset\mathbf{L}$ (column-wise or row-wise). Next, the reliabilities of each bit in the component codeword are updated. In what follows, we briefly describe this procedure in application for the single component codeword.

Chase decoding involves computing the hard decision $\mathbf{d}$ from $\bm{l}$
\begin{equation}
    \mathbf{d} = \text{bin}\left(\text{sign}\left(\bm{l}\right)\right),
\end{equation}
where $\text{sign}(\bm{l})$ denotes the sign of $\bm{l}$, and $\text{bin}(\cdot)$ is the inverse function $\tau(\cdot)$.

This step is followed by selecting the $p$ least reliable positions. The test vector set $\mathcal{T},\:|\mathcal{T}|=2^p$ consists of all binary combinations over the $p$ least reliable positions of $\mathbf{d}$. Once $\mathcal{T}$ is constructed, the Berlekamp-Massey decoding algorithm~\cite{Berlekamp1968} is applied to each test vector. This yields a candidate set of unique valid codewords $\mathcal{M},\:|\mathcal{M}|=g\leq2^p$.

If $\mathcal{M}$ is empty, the extrinsic vector is set to zero $(\mathbf{w} = \mathbf{0})$, and the component message remains unchanged, as defined in eqs.~(\ref{eq:col-update})--(\ref{eq:row-update}). If $\mathcal{M}$ is not empty, reliabilities $\mathbf{r}$ are computed according to the algorithm defined in~\cite[Sec.~5]{Pyndiah1998}. While we do not describe this algorithm in detail, we note that soft decisions $\mathbf{r}_j$ are computed for positions $\mathbf{d}_j$ covered by at least two competing codewords $\mathbf{m}^{+1(j)},\:\: \mathbf{m}^{-1(j)}\subset\mathcal{M}$. And the extrinsics for such positions are computed as
\begin{equation}
    \mathbf{w}_j = \mathbf{r}_j - \bm{l}_j
\end{equation}
Extrinsics of positions in $\mathbf{d}$ without competing candidate codewords are assigned a predefined value $\bm{\beta}_t$
\begin{equation}
    \mathbf{w}_j=\bm{\beta}_t\cdot\tau(\mathbf{d}_j).
\end{equation}
Here $t$ denotes the index of decoding half-iteration, and $\tau(\cdot)$ is the BPSK mapping function defined in eq.~(\ref{eq:modulation}). 

\section{Neural Rollback Decoding}
\label{sec:neural-rollback}

\subsection{Rollback decoding}
\label{sec:rollback}
We assume that an erroneous extrinsic information update occurs when the constructed candidate codeword set $\mathcal{M}$ does not include the originally transmitted component codeword, potentially introducing errors in the soft message update. The rollback mechanism aims to improve iterative decoding performance by preventing such erroneous updates in the Chase-Pyndiah soft-output decoding scheme. It operates as an independent block between the Chase decoding stage (candidate codeword set construction) and the Pyndiah stage (soft-output calculation). We suppose that by leveraging the composition of the candidate set $\mathcal{M}$, generated by the Chase algorithm, and the input reliability values $\bm{l}$, one can define the mechanism responsible for decision whether calculating the extrinsic information $\mathbf{w}$ and updating the component message is beneficial. If the update is likely to introduce an error, it is prevented by setting $\mathbf{w} = \mathbf{0}$. 

\begin{figure}[t]
    \centering
    \includegraphics{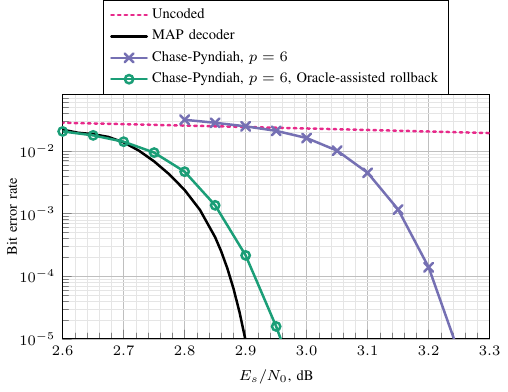}
    \caption{TPC based on $(256, 239)$ eBCH component codes. Decoding is performed for 4 full iterations ($N_T=4$).}
    \label{fig:oracle-rollback-ber}
\end{figure}

Our hypothesis that erroneous extrinsic updates occur when the candidate set $\mathcal{M}$ does not include the initially transmitted codeword can be validated numerically. In \figurename~\ref{fig:oracle-rollback-ber}, we examine the decoding performance of a TPC constructed from two $(256, 239)$ eBCH component codes over 4 full decoding iterations. We employ MAP decoding for the component codes, along with classical Chase-Pyndiah decoding (with parameter $p=6$) and \textit{Oracle} rollback decoding. The Oracle rollback mechanism operates on the a priori information of the transmitted component codeword, discarding extrinsic updates when the correct codeword is absent from the candidate set $\mathcal{M}$. In our simulations, Chase-Pyndiah algorithm parameters $\bm{\alpha}$ and $\bm{\beta}$ are not optimized and set to the values specified in the original paper~\cite{Pyndiah1998}
\begin{equation*}
    \bm{\alpha}=[0.2, 0.3, 0.5, 0.7, 0.9. 1.0, 1.0, 1.0],
\end{equation*}
\begin{equation*}
    \bm{\beta}=[0.2, 0.4, 0.6, 0.8, 1.0. 1.0, 1.0, 1.0].
\end{equation*}

The results demonstrate that Oracle rollback decoding attains MAP performance, supporting our hypothesis that the rollback mechanism in Chase-Pyndiah decoding may potentially enhance the communication system's performance.

\subsection{Data Preprocessing}
\label{sec:preprocessing}

\begin{figure}[t]
\centering
\includegraphics[width=\linewidth]{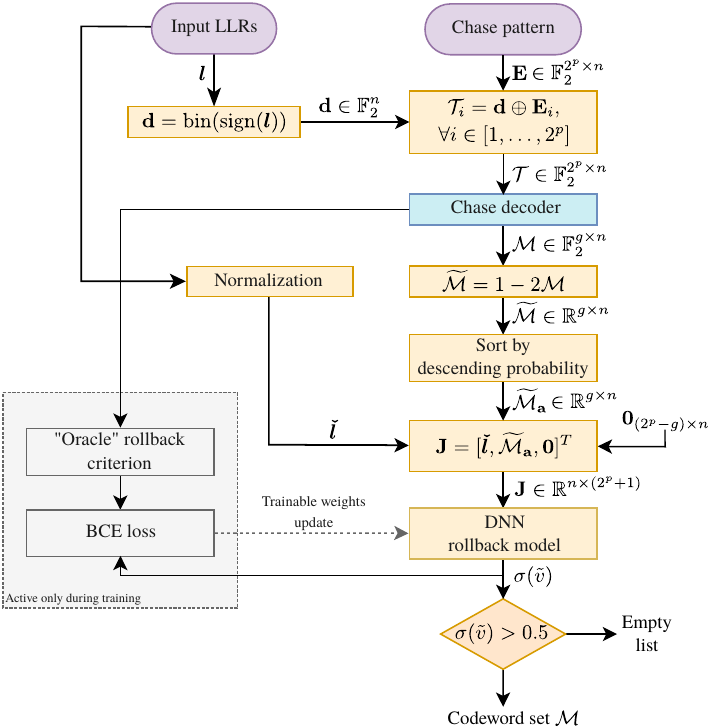}
\caption{Data processing and training pipeline.}
\label{fig:preprocessing}
\end{figure}

In this paper, we propose to use a neural network to determine whether a rollback should be performed based on the information from the Chase-Pyndiah decoder input $\bm{l}$ and the candidate codeword set $\mathcal{M}$. The data preprocessing and training scheme is summarized in the \figurename~\ref{fig:preprocessing} and described in what follows.

At first, the hard decision $\mathbf{d}$ over the decoder input $\bm{l}$ is calculated. Then, the test vector set $\mathcal{T}$ is constructed by taking the modulo-2 sum of the vector $\mathbf{d}$ and the Chase error pattern $\mathbf{E}_i,\:\:\forall i\in [1,\dots,2^p]$, which defines the positions of bits to be flipped
\begin{equation}
    \mathcal{T}_i = \mathbf{d} \oplus \mathbf{E}_i, \:\:\forall i\in [1,\dots, 2^p].
\end{equation}

In this work, we do not utilize all possible combinations of ones and zeros for the $p$ least reliable positions to construct $\mathbf{E}$. Instead, we use the landslide algorithm to construct the error patterns. This algorithm, proposed in~\cite{Duffy2022}, has demonstrated better performance in our experiments compared to the classical error pattern construction method for a fixed value of $p$ in Chase-Pyndiah decoding\footnote{This paper omits comparison with ordered reliability bits guessing random additive noise decoding (ORBGRAND), which uses the same error pattern construction method, because our experiments indicate that constructing the list of candidate codewords requires an infeasibly large number of queries. While we believe there is potential for improving this approach, exploring such enhancements is beyond the scope of this paper.}.

In the next step, Chase decoding is applied, as described in Section~\ref{sec:chase-pyndiah-decoding}, generating a set of $\mathcal{M}, \:\:|\mathcal{M}| = g$ candidate codewords. Then, BPSK modulation is applied to this set, yielding
\begin{equation}
    \mathcal{\widetilde{M}} = \tau(\mathcal{M}).
\end{equation}

Next, the correlation to the received message is computed using the inner product between the received message $\bm{l}$ and each of the BPSK-modulated codewords from $\mathcal{\widetilde{M}}$. Specifically, for each codeword $\mathcal{\widetilde{M}}_i$, the correlation is given by\footnote{The motivation for considering such a correlation stems from the well-known fact (see \cite{Feldman2005}) that maximum-likelihood decoding is equivalent to the task of minimizing the inner product of the received log-likelihood ratios $\bm{l}$ and the binary codeword $\mathbf{c}$ over all codewords in the code $\mathcal{C}$: $\hat{\mathbf{c}} = \argmin\limits_{\mathbf{c} \in \mathcal{C}} \langle \mathbf{l}, \mathbf{c} \rangle.$ The correlation considered here is equivalent because a BPSK-modulated signal, $x_{i} \in \left\{+1,-1 \right\}$, corresponds to the binary codeword symbol $c_{i} \in \left\{0,1\right\}$ via the relationship $x_{i} = 1 - 2c_{i}$ (see eq.~(\ref{eq:modulation})).}
\begin{equation}
    \mathbf{a}_i = <\bm{l}, \mathcal{\widetilde{M}}_i>,\:\:\forall{i}\in[1,\dots,g].
\end{equation}
For further description convenience, we sort the correlations
\begin{equation}
    \mathbf{a}_{(1)}\geq\mathbf{a}_{(2)} \geq \dots \geq\mathbf{a}_{(g)}
\end{equation}
and denote the sorted vector by $\mathbf{a} = [\mathbf{a}_{(1)}, \ldots, \mathbf{a}_{(g)}]$.

The input of the neural network, $\mathbf{J}\in\mathbb{R}^{n \times (2^{p} + 1)}$, is constructed by concatenating the normalized input message $\bm{\check{l}}$, a candidate set $\mathcal{\widetilde{M}}_\mathbf{a}$, sorted in descending order based on $\mathbf{a}$, and a zero matrix $\mathbf{0}_{(2^p-g)\times n}$, i.e., 

\begin{equation}
    \mathbf{J} = [\bm{\check{l}}, \mathcal{\widetilde{M}}_\mathbf{a}, \mathbf{0}]^T,
\end{equation}
where $[\cdot,\cdot,\cdot]$ denotes row-wise concatenation. The normalized input message is defined as 
\begin{equation}
    \bm{\check{l}} = \sqrt{n}\frac{\bm{l}}{\lVert\bm{l}\rVert_2}.
\end{equation}
Here, the normalization of $\bm{l}$ is applied to improve training stability by ensuring a consistent scale across each non-zero column of the model input\footnote{The normalization of $\bm{l}$ should not be confused with that of $\mathbf{\Gamma}$. The matrix $\mathbf{\Gamma}$ is normalized to align with the values of the parameters $\bm{\beta}$, while $\bm{l}$ is normalized to maintain a consistent scale for the component message and BPSK-modulated candidate codewords on the input of the model.}. The inclusion of the zero-padding matrix $\mathbf{0}$ guarantees that the input to the model maintains a fixed shape.

\subsection{Neural Network Architecture}
\begin{figure}[t]
\centering
\includegraphics[width=0.75\linewidth]{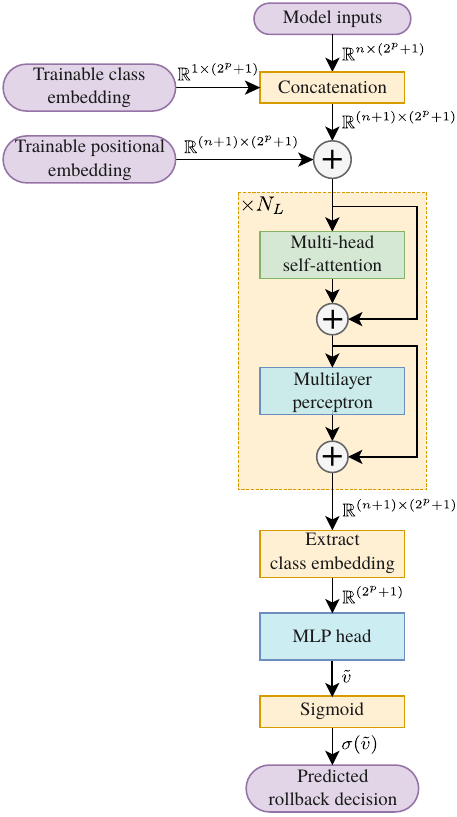}
\caption{Rollback transformer architecture.}
\label{fig:architecture}
\end{figure}

Inspired by state-of-the-art neural network architectures, we utilize transformer\cite{Vaswani2017} to determine if the rollback action is required. The architecture is depicted in \figurename~\ref{fig:architecture}. Specifically, we use the encoder part of the transformer in a manner similar to its application in classification tasks in visual transformers (ViT)~\cite{Dosovitskiy2021}. The matrix $\mathbf{J}$ is utilized as the model's input. Throughout the following description, we treat the code length dimension $n$ as the token dimension. As the first step, a trainable class embedding $\bm{\theta}_c\in\mathbb{R}^{1\times (2^p+1)}$ is concatenated to the model's input $\mathbf{J}\in\mathbb{R}^{n\times (2^p+1)}$
\begin{equation}
\label{eq:model-input}
    \mathbf{J}_c = [\bm{\theta}_c, \mathbf{J}].
\end{equation}

Next, a trainable positional embedding $\mathbf{\Theta}_p$ is added to $\mathbf{J}_c$
\begin{equation}
    \mathbf{J}_p = \mathbf{J}_c + \mathbf{\Theta}_p, \quad \mathbf{J}_p\in\mathbb{R}^{(n+1)\times (2^p+1)}.
\end{equation}

Then, a series of $N_L$ transformer blocks, each consisting of multi-head self-attention (MHSA) and a multilayer perceptron (MLP), is applied.

Since the rollback decision can be treated as the binary classification task, the prediction, as in~\cite{Dosovitskiy2021}, is based on the class label. Thus, only the first row of the transformer output $\mathbf{J}_{\text{out}}$ is fed into the final MLP-head layer, which reduces its dimension from $2^p+1$ to a scalar value. The rollback decision is determined from the probability obtained by applying a sigmoid function to the MLP-head output $\tilde{v}$. If $\sigma(\tilde{v})>0.5$, the candidate codeword set $\mathcal{M}$ is passed to the Pyndiah soft-output approximation algorithm (see~Section~\ref{sec:chase-pyndiah-decoding}). Otherwise, rollback is triggered, and the zero extrinsic information vector ($\mathbf{w} = \mathbf{0}$) is returned.

The training of the model is performed in a sequential manner on randomly generated binary TPC codewords. Let us consider the training procedure at the half-iteration $t$. The model input is produced by applying previously trained neural network-based rollback decoders for $t-1$ half-iterations, followed by an Oracle-based rollback decoder at the half-iteration $t$\footnote{At the first half-iteration, only the Oracle decoder is used to produce inputs for the trainable model.}. During this half-iteration, the Oracle rollback decoder produces vectors $\bm{l}$, candidate codeword sets $\mathcal{M}$, and hard rollback decisions $v\in{\{0,1\}}$. The neural network inputs are then constructed from $\bm{l}$ and $\mathcal{M}$, as defined in eq.~(\ref{eq:model-input}). The loss is computed using the Oracle decisions $v$ and the neural network output probabilities $\sigma(\tilde{v})$. The binary cross-entropy is utilized as the loss function
\begin{equation}
\label{eq:bce-loss}
    \mathcal{L}_{BCE}(\tilde{v},v) = -\Big[v\log\sigma(\tilde{v})+ 
    (1-v)\log(1-\sigma(\tilde{v}))\Big].
\end{equation}
Once the model is trained at half-iteration $t$, its weights are frozen, and it is used in the decoding scheme to generate outputs for the next half-iteration, $t+1$. This process continues until all models for $N_T$ iterations are trained\footnote{In the described setup, separate trainable weights are optimized for each half-iteration. However, it may be possible to perform rollback decoding with a single set of weights. We leave this question for future investigation.}.
\section{Experimental Setup and Simulation Results}
\label{sec:experiments-results}

We consider a TPC scheme, described in Section~\ref{sec:system-model}, based on the $(256,239)$ eBCH component codes. The parameters of the rollback NN models are summarized in Table~\ref{tab:model-parameters}. Each model was trained for $2500$ epochs. The initial learning rate was set to $10^{-4}$ and was subsequently reduced to $10^{-6}$ using the ``reduce on plateau'' scheduler. For the considered code, AWGN was generated with a variance sampled from a uniform distribution, corresponding to an SNR range of $[2.95, 3.05]$~dB.

\begin{table}
    \centering
    \caption{Neural rollback transformer model parameters}
    \label{tab:model-parameters}
    \begin{tabular}{|c|c|}
    \hline
    Parameter type & Parameter value \\ \hline \hline
    Hidden dimension & $2^p+1=65$\\ \hline
    Transformer encoder depth & 2 \\ \hline
    Number of MHSA heads & 4 \\ \hline
    MHSA head hidden dimension & 256 \\ \hline
    MLP hidden dimension & 256 \\ \hline
    \end{tabular}
\end{table}

Here, we compare the performance of the proposed decoding method not only with those described in Section~\ref{sec:neural-rollback} but also with simple rollback criteria inspired by the analysis conducted in~\cite{Forney1968}. The first criterion, denoted as \mbox{\textit{Top-1}}, is a threshold-based approach that discards the extrinsic information update if the maximum correlation $\varphi_1 = \mathbf{a}_{(1)}$, falls below a predefined threshold $\mu_1$. The motivation behind this criterion is that if the most probable codeword from the candidate set $\mathcal{M}$ deviates significantly from the initial codeword, the decoding procedure is likely to introduce an error. The second criterion, denoted as \mbox{\textit{Top-2}}, is based on the difference between the two highest correlation values $\varphi_2 = \mathbf{a}_{(1)} - \mathbf{a}_{(2)}$. If $\varphi_2$ exceeds a predefined threshold $\mu_2$, the list of codewords is likely to contain a reliable decision and is returned to the Pyndiah algorithm. Otherwise, an empty list is returned.

To determine the decision thresholds, we use Monte Carlo simulations combined with the Nelder–Mead optimization method~\cite{Nelder1965} to find the optimal thresholds $[\mu_1^{(1)},\dots,\mu_1^{(2N_T)}]$ and $[\mu_2^{(1)},\dots,\mu_2^{(2N_T)}]$, for each decoding half-iteration.

Also, we consider the performance of MAP-based rollback decoding, which operates as follows. If the candidate codeword set $\mathcal{M}$ includes the hard-decision output of the MAP decoder, then $\mathcal{M}$ is passed to the Pyndiah soft-output calculation algorithm. Otherwise, no extrinsic information update is performed.

\begin{figure}[ht]
    \centering
    \includegraphics{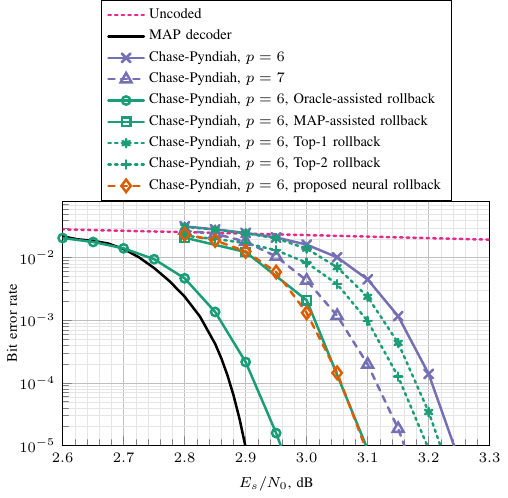}
    \caption{TPC based on $(256, 239)$ eBCH component codes. Decoding is performed for 4 full iterations ($N_T=4$).}
    \label{fig:neural-rollback-ber}
\end{figure}

The performance of the described algorithms is shown in \figurename~\ref{fig:neural-rollback-ber}.  From the figure, we observe that the \mbox{\textit{Top-1}} and \mbox{\textit{Top-2}} rollback methods improve decoding performance but remain significantly below the MAP-assisted and proposed neural rollback methods. 
The proposed neural rollback decoding achieves performance comparable to MAP-assisted rollback, outperforms Chase-Pyndiah decoding with $p=7$, and surpasses Chase-Pyndiah decoding with $p=6$ by approximately $0.145$~dB on BER level of $10^{-4}$. Furthermore, the proposed method attains roughly half the performance gain of the Oracle-based rollback Chase-Pyndiah $p=6$ decoding scheme.

\section{Conclusion}
\label{sec:conclusion}

This paper proposes a novel iterative rollback decoding method based on a transformer neural network. The approach effectively detects and prevents erroneous extrinsic information updates, addressing the limitations of conventional Chase-Pyndiah decoding. Evaluations on TPC with $(256, 239)$ eBCH component codes demonstrate that the proposed method improves BER, achieving a gain of approximately $0.145$~dB for Chase-Pyndiah $p=6$ over four full decoding iterations. This performance surpasses conventional Chase $p=7$ decoding, underscoring the potential of neural network-assisted mechanisms in enhancing soft-decision decoding for practical communication systems. Furthermore, the proposed method can be implemented as a standalone module operating between the Chase and Pyndiah stages without modifying the classical algorithms.

Future research directions include refining the training procedure and model architecture to enhance generalization across all decoding iterations with a single set of trainable weights. Additionally, further exploration is needed to improve the model’s ability to generalize across different Chase-Pyndiah parameters $p$, as well as various code types and lengths. Another important avenue for future work is reducing model complexity to enhance efficiency for real-world hardware implementation.

\bibliographystyle{IEEEtran}
\bibliography{refs}

\end{document}